\documentstyle[12pt]{article}
\begin{document}

\def\star{{\displaystyle *}}
\def\be{\begin{equation}}
\def\ee{\end{equation}}

\begin{center}
{\large {\bf Self-consistent fluid model for a quantum electron gas}}
\vskip 5mm

{G. Manfredi\footnote{Giovanni.Manfredi@lpmi.uhp-nancy.fr}, 
F.~Haas\footnote{ferhaas@lncc.br}} \\ Laboratoire de 
Physique des Milieux Ionis\'es, Universit\'e Henri Poincar\'e, \\
BP 239, 54506 Vandoeuvre-les-Nancy, France 
\vskip 5mm
\end{center}
\begin{abstract}
\noindent
It is shown that, for a large class of statistical mixtures, 
the Wigner-Poisson (or Hartree) system can be reduced to an 
effective Schr\"odinger-Poisson system,
in which the Schr\"odinger equation contains a new nonlinearity.
For the case of a zero-temperature one-dimensional electron gas, 
this additional nonlinearity is of the form
$\vline\Psi\vline~^4$.
In the long wavelength limit, the results obtained
from the effective 
Schr\"odinger-Poisson system are in agreement with those of the Wigner-Poisson system.
The reduced model is further used to describe the stationary states of a 
quantum electron gas and the two-stream instability.
\end{abstract}

\section{Introduction}

Understanding the dynamics of a quantum electron gas is an important issue 
for a variety of physical systems, such as ordinary metals, semiconductors, and 
even astrophysical systems under extreme conditions (e.g. white dwarfs).
Although some level of understanding can be achieved by considering independent 
electrons, a more accurate description requires the use of self-consistent models,
where electron-electron interactions are taken into account. As the treatment of
the full $N$-body problem is clearly out of reach, mean-field models are usually adopted,
of which the Hartree and Hartree-Fock models are standard examples \cite{Ashcroft}.
In the Hartree approximation, each electron is described by a one-particle
wave function (obeying Schr\"odinger's equation), and the electrostatic 
force acting on it results from Poisson's equation. (Fock's correction accounts for
the parity of the $N$-particle wave function for an ensemble of Fermions, but this
correction will not be considered in this paper).

The Hartree model can be written in a more compact, but strictly equivalent, form by
making use of Wigner functions. The Wigner representation \cite{Wigner} 
is a useful tool to express quantum mechanics in a phase space formalism 
(for reviews see \cite{Tatarskii}). 
In this representation, a quantum state (either pure or mixed) 
is described by a Wigner function (i.e. a function of the phase
space variables), and the Wigner equation 
provides an evolution equation which is similar to the
Vlasov equation, well-known from classical plasma physics.
We note that, although the Wigner distribution satisfies most of the
standard properties of probability distributions,
it cannot be regarded as such, since it may take negative values. 
The resulting self-consistent model is called the Wigner-Poisson (WP) system, and
has been extensively used in the study of quantum transport 
\cite{Kluksdahl}-\cite{Drummond}. 

Despite its considerable interest, the WP formulation presents
some intrinsic drawbacks : (a) it is a nonlocal, integro-differential system;
and (b) its numerical treatment requires the discretisation of 
the whole phase space.
Moreover, as is often the case with kinetic models, the WP system
gives more information than one is really interested in. 
For these reasons, it would be useful to obtain an accurate reduced model
which, though not providing the same detailed information as the kinetic
WP system, is still able to reproduce the main 
features of the physical system under consideration. 

After discussing the general validity of the WP model, 
we will derive an effective Schr\"odinger-Poisson (SP) 
system, which, in 
an appropriate limit, reproduces the results of the kinetic WP formulation. 
A similar result was recently obtained in the mathematical literature
\cite{Gardner, Gasser}, although its physical implications have not been 
fully analysed. 
In this effective 
SP model, the Schr\"odinger equation is nonlinear, as
it includes an effective potential depending on the modulus of the wavefunction. 
The exact form of this 
effective potential depends on the specific physical system being studied. 
In order to obtain the effective SP system, we will first derive
a system of reduced `fluid' equations by taking moments of the WP system. 
It will be shown that the pressure 
term appearing in the fluid equations can be decomposed into a classical 
and a quantum part. With some reasonable hyphoteses on the
pressure term, the fluid system can be closed. 
Finally, the effective SP system will be applied to several physical problems,
including linear wave propagation, nonlinear stationary solutions, and the 
two-stream instability.


\section{Coupling parameter for a quantum plasma}

A classical plasma can be said to be collisionless (`ideal') when long-range 
self-consistent interactions (described by the Poisson equation)
dominate over short-range two-particle interactions (collisions). 
This happens when the potential energy
of two electrons separated by an average interparticle distance is small compared to
the average kinetic energy. The potential energy is estimated as 
$E_{pot} = e^2 n_0^{1/3} / \varepsilon_0$, while the average kinetic energy
is simply given by the temperature $T$ (measured in energy units).
Here $-e$ is the electron charge, $\varepsilon_0$ the dielectric constant in
vacuum, and $n_0$ the equilibrium particle density.
One defines therefore a classical coupling parameter
\be
\label{gc}
\Gamma_C = \frac{E_{pot}}{E_{kin}} = \frac{e^2 n_0^{1/3}}{\varepsilon_0 T}~,
\ee
such that the collisionless 
approximation is valid when $\Gamma_C \ll 1$. 
The classical coupling parameter can be written in a different way, by introducing 
the plasma frequency, the thermal velocity and the Debye length
\be
\omega_p = \left(\frac{n_{0}e^{2}}{m\varepsilon_{0}}\right)^{1/2},~~
v_T = \left(\frac{T}{m}\right)^{1/2},~~
\lambda_D = \frac{v_T}{\omega_p}~,
\ee

which are typical inverse time, velocity and length scales for a collisionless plasma.
With these definitions, the coupling parameter can be expressed as
\be
\label{gc2}
\Gamma_C^{3/2} = \frac{1}{n_0 \lambda_D^3}~,
\ee
which is the inverse of the number of electrons contained in a Debye volume.
When the condition $\Gamma_C \ll 1$ is satisfied, two-body correlations (collisions)
can be neglected, and the $N$-particle Liouville equation can be reduced, via a
BBGKY hierarchy, to the one-particle Vlasov equation. The Vlasov-Poisson system is
therefore the standard model to describe classical electrostatic plasmas in the
collisionless approximation.

Similarly, it is possible to define a quantum coupling parameter
$\Gamma_Q$. Let us consider the case of a completely degenerate electron gas. Now
the average kinetic energy is given by the
Fermi temperature $T_F \sim \hbar^2 n_0^{2/3}/m$ (we neglect irrelevant dimensionless
constants), so that the quantum coupling parameter becomes

\be
\label{gq}
\Gamma_Q = \frac{E_{pot}}{T_F} = \frac{e^2 m}{\hbar^2 \varepsilon_0 n_0^{1/3}}~.
\ee
Notice that, according to Eq. (\ref{gq}), a quantum electron
gas is more ideal at higher densities. 
Using the Fermi velocity $v_F = \sqrt{T_F/m}$, one can define a typical length scale for the 
quantum plasma 
\be
\label{lf}
\lambda_F = \frac{v_F}{\omega_p}~.
\ee
The quantum coupling parameter can thus be expressed as the inverse of the
number of electrons contained in a Fermi volume :
\be
\label{gq2}
\Gamma_Q^{3/2} = \frac{1}{n_0 \lambda_F^3}~.
\ee
Finally, another expression for the coupling parameter is the following
\be
\label{gq3}
\Gamma_Q^{1/2} = \frac{\hbar \omega_p}{T_F}~,
\ee
which is valid for any number of dimensions.

The quantum electron gas is collisionless when $\Gamma_Q \ll 1$.
In this case, the quantum $N$-body problem can be reduced to a one-particle
Wigner equation. The Wigner-Poisson system is therefore capable of describing 
a quantum electrostatic plasma in the collisionless approximation.

The previous results were derived in the limiting cases $T \gg T_F$ (classical) and 
$T \ll T_F$ (quantum degenerate). For intermediate temperatures, simple expressions 
for the coupling parameters are not available, but one must expect a smooth transition
between the two regimes.

For electrons in metal, we have typically 
\be
n_0 \simeq 10^{29} ~{\rm m^{-3}},~~~
v_F \simeq 10^6 ~{\rm m s^{-1}},~~~
\omega_p \simeq 10^{16} ~{\rm s^{-1}} ,~~~
\lambda_F \simeq 10^{-10}~ {\rm m}~.
\ee
These values yield a quantum coupling parameter of order unity. Allowing for
the dimensionless constants we have neglected and the different properties of metals,
we realise that $\Gamma_Q$ can be both smaller and larger than unity for typical metallic
electrons.

The above values seem to indicate that,
as $\Gamma_Q \simeq 1$, electron-electron (e-e) collisions cannot be
neglected for metals. If that was the case, one should abandon one-particle models
such as the Wigner or Hartree equations, and resort to the full $N$-body problem.
This is hardly a feasible task. Fortunately, however, the exclusion principle comes 
to rescue by reducing the collision rate quite dramatically in most cases of 
interest \cite{Ashcroft}. This occurs when the electron distribution is close to 
the Fermi-Dirac equilibrium at relatively low temperatures. The fundamental point
is that, when all lower levels are occupied, the exclusion principle forbids
a vast number of transitions that would otherwise be possible.
In particular, at strictly zero temperature, all electrons have energies below
$T_F$, and no transition is possible, simply because there are no available 
states for the electrons to occupy.
At moderate temperatures, only electrons wihin a shell of thickness $T$
about the Fermi surface can undergo collisions. The e-e collision rate
(inverse of the lifetime $\tau_{ee}$) for such electrons is proportional to
$T/\hbar$ (this is a form of the uncertainty principle, energy $\times$ time = const.).
The {\it average} collision rate is obtained by multiplying the previous expression by
the fraction of electrons present in the shell of thickness $T$
about the Fermi surface, which is $\sim T/T_F$. One obtains
\be
\label{tauee}
\frac{1}{\tau_{ee}} \sim \frac{1}{\hbar} ~\frac{T^2}{T_F}~.
\ee
At room temperature, $\tau_{ee} \simeq 10^{-10}~s$, which is much larger than
the typical collisionless time scale $\tau_p = \omega_p^{-1} \simeq 10^{-16}~s$.
Therefore, for times smaller than $\tau_{ee}$ the effect of e-e collisions
can be safely neglected. In addition, it turns out that the typical 
relaxation time scale is $\tau_{r} \simeq 10^{-14}~s$, which is again
significantly larger than $\tau_p$. In summary, the ordering
\be
\label{order}
\tau_p \ll \tau_{r} \ll \tau_{ee}~,
\ee 
implies that a collisionless (Wigner) model is appropriate for relatively 
short time scales.


\section{Derivation of the fluid model} 

In one spatial dimension, the Wigner-Poisson system \cite{Wigner,Tatarskii} reads as
\begin{eqnarray}
\frac{\partial{f}}{\partial{t}} &+& v\frac{\partial{f}}{\partial{x}}+\frac{iem}{2\pi\hbar}
\int\int{d\lambda}~{dv'}e^{im(v-v')\lambda} \times \nonumber \\
\label{wig}
& \times &
\left[\phi\left(x+\frac{\lambda \hbar}{2}\right)-
\phi\left(x-\frac{\lambda \hbar}{2}\right)\right]f(x,v',t) = 0 \,,  \\
\label{pois}
\frac{\partial^{2}\phi}{\partial\,x^2} &=& \frac{e}{\varepsilon_0}\left(\int f\,dv - 
n_{0} \right) ~,
\end{eqnarray}
where $f(x,v,t)$ is the Wigner distribution function, 
$\phi(x,t)$ the electrostatic potential, $-e$ and 
$m$ the electron charge and mass, $\varepsilon_0$ the vacuum dieletric constant and 
$n_0$ a background ionic charge. 
Notice that the one-particle Wigner function used here actually represents
an $N$-particle system. Indeed, the above Wigner-Poisson system can be derived
from the full $N$-body problem via a BBGKY hierarchy, neglecting two-body correlations
and only keeping the mean Coulomb field \cite{balescu}.
Further, it is easy to see that, in the limit $\hbar \to 0$, 
one recovers the familiar Vlasov-Poisson system for classical collisionless plasmas.
For simplicity of notation, only one-dimensional problems will be
treated in the rest of this paper, but the results can be readily 
extended to higher dimensions.

In order to derive a fluid model, we take moments of Eq. (\ref{wig}) by integrating
over velocity space. Introducing the standard definitions of density,
mean velocity and pressure 
\begin{equation}
\label{fluid}
n(x,t) = \int f\,dv \,, \quad u(x,t) = \frac{1}{n}\int fv\,dv \,, \quad P(x,t) = 
m\left(\int fv^{2}dv - nu^{2}\right) \, ,
\end{equation}
it is obtained
\begin{eqnarray}
\label{cont}
\frac{\partial\,n}{\partial\,t} + 
\frac{\partial\,(nu)}{\partial\,x} &=& 0 \,, \\ 
\label{force}
\frac{\partial\,u}{\partial\,t} + u\frac{\partial\,u}{\partial\,x} &=& 
\frac{e}{m}\frac{\partial\,\phi}{\partial\,x} - \frac{1}{mn}
\frac{\partial\,P}{\partial\,x} \,.
\end{eqnarray}

We immediately notice that Eqs. (\ref{cont})-(\ref{force}) 
do not differ from the ordinary evolution equations for a classical fluid.
This may seem surprising, but in the following it will appear that the quantum
nature of this system is in fact hidden in the pressure term.

The pressure term may be decomposed into a classical and a quantum part. 
This can be shown as follows. The Wigner distribution for a quantum 
mixture of states $\psi_{\alpha}(x,t)$, each characterized by an occupation probability 
$p_\alpha$, is written as
\begin{equation}
\label{wigfun}
f(x,v,t) =  \frac{m}{2\pi\hbar}\sum_{\alpha}p_{\alpha}\int d\lambda~
\psi^{*}_{\alpha}\left(x + \frac{\lambda}{2},t\right)
\psi_{\alpha}\left(x - \frac{\lambda}{2},t\right) e^{imv\lambda/\hbar}\,,
\end{equation}
where the sum extends over all possible states. The numbers $p_\alpha$, representing
probabilities, satisfy the relations $p_\alpha \geq 0$, $\sum_{\alpha}p_\alpha = 1$. 
Using the the previous expression, one can compute the pressure. After some
algebra, one obtains
\[
P = \frac{\hbar^2}{4m}  \sum_{\alpha} p_\alpha\left( 2 
\left|\frac{\partial \psi_\alpha}{\partial x}\right|^2 - 
\psi^\star_\alpha \frac{\partial^2 \psi_\alpha}{\partial x^2} -
\psi_\alpha \frac{\partial^2 \psi^\star_\alpha}{\partial x^2} \right) 
\]
\be
+ ~\frac{\hbar^2}{4mn} \left[ \sum_{\alpha} p_\alpha\left(
\psi^\star_\alpha \frac{\partial \psi_\alpha}{\partial x} -
\psi_\alpha \frac{\partial \psi^\star_\alpha}{\partial x} \right)
\right]^2 ~.
\ee

If we represent each state as 
\begin{equation}
\label{psi}
\psi_{\alpha}(x,t) = A_{\alpha}(x,t)\exp{(iS_{\alpha}(x,t)/\hbar)}~,
\end{equation}
where $A_\alpha$ (amplitude) and $S_\alpha$ (phase) are real functions, 
we obtain $P = P^{C} + P^{Q}$, 
where the classical $P^C$ and quantum $P^Q$ parts of the pressure are
\begin{eqnarray}
\label{pc}
P^C &=& \frac{1}{2mn}\sum_{\alpha,\beta}p_{\alpha}p_{\beta}A_{\alpha}^{2}A_{\beta}^{2}
\left(\frac{\partial S_\alpha}{\partial x}- \frac{\partial S_\beta}{\partial x}\right)^2 \,,\\
\label{pq}
P^Q &=& \frac{\hbar^2}{2m}\sum_{\alpha}p_{\alpha}\left[\left(\frac{
\partial\,A_\alpha}{\partial\,x}\right)^2 - A_{\alpha}\frac{\partial^{2}A_{\alpha}}
{\partial\,x^2}\right] ~.
\end{eqnarray}
Notice that $P^Q$ only depends on the amplitudes $A_\alpha$, 
and that for a pure state only $P^Q$ survives. It can be easily shown 
that $P^C$ represents the 
standard pressure, resulting from the dispersion of 
velocities. 
To prove this, one has to remember that the phases $S_\alpha$ 
are related to the mean velocity $u_\alpha$ of each wavefunction through
the relation $m u_\alpha = \partial S_\alpha/\partial x$ [the $u_\alpha$'s
should not be mistaken with the {\it global} mean velocity $u$ defined in
Eq. (\ref{fluid})]. Thus, by expanding Eq. (\ref{pc}), one obtains after 
some algebra
\begin{equation}
\label{pc2}
P^C = mn \left[\frac{\sum_{\alpha} p_\alpha A^2_\alpha u^2_\alpha}{n}
- \left(\frac{\sum_\alpha p_\alpha A^2_\alpha u_\alpha}{n}\right)^2
\right] ~.
\end{equation}
With an appropriate definition of averages, we can rewrite the above equation 
as: $P^C = mn (\langle u_\alpha^2 \rangle - \langle u_\alpha \rangle^2)$,
which is the standard expression for the pressure.
The contribution $P^Q$, on the other hand, is 
a purely quantum pressure, with no classical counterpart. 

In order to close the fluid system, some equation of state, 
relating $P^C$ and $P^Q$ to the density $n$, 
must be used. In this paper, we consider a statistical mixture where all the 
amplitudes are equal 
(but not constant), $A_\alpha(x) = A(x)$.
This gives, using Eqs. (\ref{fluid}) and (\ref{wigfun}), $n = A^2$. 
With the help of Eq. (\ref{pq}), the quantum pressure becomes
\begin{equation}
\label{pq2}
P^Q = \frac{\hbar^2}{2m}\left[\left(\frac{\partial}{\partial x}\sqrt{n}\right)^{2} 
- \sqrt{n}\frac{\partial^2}{\partial\,x^2}\sqrt{n}\right] \,.
\end{equation}
For the classical part of the pressure, we make the standard assumption that it is some
function of the density, $P^C = P^{C}(n)$. With these hypotheses, the force equation 
$(\ref{force})$ can be written as
\begin{equation}
\label{force1}
\frac{\partial\,u}{\partial\,t} + u\frac{\partial\,u}{\partial\,x} =
\frac{e}{m}\frac{\partial\,\phi}{\partial\,x} - 
\frac{1}{mn}\frac{\partial\,P^C}{\partial\,x} -
\frac{1}{mn}\frac{\partial\,P^Q}{\partial\,x} \,.
\end{equation}
Defining the effective potential
\begin{equation}
\label{efpot}
W(n) = \int^{n}\frac{dn'}{n'}\frac{dP^{C}(n')}{dn'} \,,
\end{equation}
and using the identity
\be
\frac{1}{mn}\frac{\partial\,P^Q}{\partial\,x} =
- \frac{\hbar^2}{2m^2}
\frac{\partial}{\partial x}\left(\frac{\partial^{2}(\sqrt{n})/\partial\,x^2}
{\sqrt{n}}\right)~,
\ee
the force equation $(\ref{force1})$ reduces to
\begin{equation}
\label{force2}
\frac{\partial\,u}{\partial\,t} + u\frac{\partial\,u}{\partial\,x} =
\frac{e}{m}\frac{\partial\,\phi}{\partial\,x} - 
\frac{1}{m}\frac{\partial\,W}{\partial\,x} +\frac{\hbar^2}{2m^2}
\frac{\partial}{\partial x}\left(\frac{\partial^{2}(\sqrt{n})/\partial\,x^2}
{\sqrt{n}}\right) \,.
\end{equation}

Now comes the crucial point : it is possible to
combine Eqs. (\ref{cont}) and (\ref{force2}) into an effective Schr\"odinger 
equation. Indeed, let us define the effective wavefunction 
\begin{equation}
\Psi = \sqrt{n(x,t)}\exp{(iS(x,t)/\hbar)} \,,
\end{equation}
with $S(x,t)$ defined according to $m u(x,t) = \partial\,S(x,t)/\partial\,x$.
We obtain that $\Psi(x,t)$ satisfies the equation
\begin{equation}
\label{nlse}
i\hbar\frac{\partial\Psi}{\partial\,t} = 
- \,\frac{\hbar^2}{2m}\frac{\partial^{2}\Psi}{\partial\,x^2} - 
e\phi\Psi + W\Psi \, .
\end{equation}
This is a nonlinear Schr\"odinger equation, as the effective potential $W$ 
depends on the wavefunction through Eq. (\ref{efpot}), where $n = |\Psi|^2$. 
Separating Eq. (\ref{nlse}) into its real and imaginary parts, we indeed find the 
continuity (\ref{cont}) and force (\ref{force2}) equations. Finally,
the complete effective SP system is composed of Eq. (\ref{nlse}) 
and the Poisson equation 
\begin{equation}
\label{po}
\frac{\partial^{2}\phi}{\partial\,x^2} = 
\frac{e}{\varepsilon_0}\left(|\Psi|^2 - n_{0}\right) \,.
\end{equation}

To summarize what we have achieved so far, we
notice that, in general, the dynamics of a statistical mixture 
must be treated with the full 
Wigner-Poisson system, or, equivalently, with a set of Schr\"odinger equations,
coupled by Poisson's equation (Hartree's model).
In the present section, we have shown that one can reduce the problem of 
quantum transport to a single nonlinear Schr\"odinger equation plus 
Poisson's equation. Also, notice that the nature of the interaction
(electrostatic in our case) is not of essential importance. The main result
is that we can reduce the (phase space) Wigner equation to a (real space)
nonlinear Schr\"odinger equation.

The two hypotheses used for this reduction are : (a) all states composing 
the mixture have
the same amplitude [which leads to Eq. (\ref{pq2}) for the quantum pressure];
and (b) the equation of state for the classical pressure is $P^C = P^{C}(n)$. 
Hypothesis (b) is the standard fluid closure, and needs no further comment.
Hypothesis (a) means that all electrons are distributed in space according
to the same probability distribution 
$n(x)=A^2(x)$. 
What distinguishes the electrons
from one another is their phase $S_\alpha$, and therefore their velocity $u_\alpha$.
This approximation can be viewed as a first step beyond
the standard homogeneous equilibrium of a Fermion gas, for which each state
is represented by a plane wave
\[
\psi_{\alpha}(x,t) = A \exp{(i m u_\alpha x/\hbar})~,
\]
with the amplitude $A$ and 
the velocity $u_\alpha$ spatially constant.
In our approximation, both the amplitude and the velocity
can be spatially modulated, although we still restrict ourselves to the case
where the amplitude is the same for all states.
This appears to be a reasonable closure assumption for 
systems that are not too far from the Fermi-Dirac equilibrium.


\section{Applications}

As a relevant example of the above theory, we consider a 
zero-temperature one-dimensional electron gas, with Fermi velocity 
$v_F$ and equilibrium density $n_0$. In this case, the classical
pressure is 
\begin{equation}
\label{pdeg}
P^C = \frac{mv_{F}^2}{3n_{0}^2}n^3 \,.
\end{equation}
(Notice that the term `classical' is somewhat inappropriate
here, as $P^C$ will contain Planck's constant through the Fermi velocity).
We also note that the Fermi velocity in one spatial 
dimension 
\be
v_F = {\pi \over 2} \frac{\hbar n_0}{m}
\ee
is proportional to $n_0$, whereas in three dimensions $v_F \propto n_0^{1/3}$.

Using Eq. (\ref{pdeg}), the effective potential defined in Eq.  
(\ref{efpot}) turns out to be 
\begin{equation}
\label{efpotdeg}
W = \frac{mv_{F}^{2}}{2n_{0}^{2}}|\Psi|^4 \,.
\end{equation}
%
Notice that the effective potential is repulsive, and tends to flatten the
electron density. This is quite natural, as $W$ derives from the pressure
$P^C$, which in turn is a manifestation of the dispersion of velocities
in a Fermion gas. When the gas is at equilibrium, $W \sim n \sim $ const., and this
term has no effect.

We also point out that a similar nonlinear Schr\"odinger 
equation with a $|\Psi|^4$-dependent potential has recently been derived in the study 
of low-dimensional Bose condensates \cite{condensed}. 
We stress, however, that such a Boson-Fermion duality only applies to
one-dimensional systems.
For a $D$-dimensional Fermion system, the classical part of the pressure has the
form $P^C \sim n^{(D+2)/D}$, so that the effective potential
becomes $W \sim n^{2/D}$ \cite{nagy}.


\subsection{Linear wave propagation}

As a first application, let us study wave propagation for the effective SP
system (\ref{nlse})-(\ref{po}) with $W$ given by Eq. (\ref{efpotdeg}). 
Linearizing around the homogeneous
equilibrium $\Psi = \sqrt{n_0}$, $e\phi = mv_{F}^{2}/2$, we obtain the following
dispersion relation (for waves with frequency $\omega$ and wave number $k$)  
\begin{equation}
\label{dispfluid}
\omega^2 = \omega_{p}^2 + k^{2}v_{F}^2 + \frac{\hbar^{2}k^4}{4m^2} \,. 
\end{equation}
For $v_F = 0$, we recover the dispersion relation of the standard SP system 
\cite{Drummond}. Equation (\ref{dispfluid}) can be written in dimensionless
units by using Eqs. (\ref{lf}) and (\ref{gq3}), which are valid both in one
and three spatial dimensions
\begin{equation}
\label{dispfluid2}
{\omega^2 \over \omega_{p}^2} = 1 + k^{2}\lambda_{F}^2 + 
\frac{k^4\lambda_{F}^4 }{4} ~\Gamma_Q \,.
\end{equation}
Note that quantum {\it mechanical} effects (dispersion of the
wave packet) are first order 
in the coupling parameter $\Gamma_Q$, whereas quantum {\it statistical}
effects (Fermi-Dirac distribution) appear at leading order. 

We want to compare this dispersion relation to the one obtained from the
complete WP system (\ref{wig})-(\ref{pois}), which, 
in the most general case, reads as \cite{Drummond, Suh}
\begin{equation}
\label{dispkin1}
1 - \frac{\omega_{p}^2}{n_{0}}\int\frac{f_{0}(v)~dv}
{(\omega - kv)^{2} - 
\hbar^{2}k^{4}/4m^{2}} = 0 \,.
\end{equation}
In our case, $f_0(v)$ is
given by the Fermi-Dirac distribution for a
zero-temperature one-dimensional electron gas at equilibrium, i.e.
$f_{0}(v) = n_{0}/2v_{F}$ if $|v| < v_{F}$ and $f_{0}(v) = 0$ if $|v| > v_{F}$.
Substituting into Eq. (\ref{dispkin1}), one obtains (without any further
approximation)
\begin{equation}
\label{dispkin2}
{\omega^2 \over \omega_p^{2}} = \frac{\Omega^2}{\omega_{p}^2} 
\coth\left(\frac{\Omega^2}{\omega_{p}^2}\right) + 
k^{2}\lambda_{F}^2 + \frac{k^4\lambda_{F}^4}{4} ~\Gamma_Q  \,, 
\end{equation}
where
\begin{equation}
\label{def}
{\Omega^2 \over \omega_p^{2}} = \frac{\hbar\,k^{3}v_F}{m \omega_p^2} 
= k^3 \lambda_{F}^3~ \Gamma_Q^{1/2} \,.
\end{equation}

Now we expand the first term on the right-hand side of Eq. (\ref{dispkin2})
in the long wavelength (fluid) limit $\Omega \ll \omega_p$. Using the
expansion $x \coth(x) = 1 + x^2/3 - x^4/45 + \dots$, one obtains

\begin{equation}
\label{dispkin3}
{\omega^2 \over \omega_p^{2}} = 1 +  k^{2}\lambda_{F}^2 + 
\left(\frac{k^4 \lambda_{F}^4}{4} + \frac{k^6 \lambda_{F}^6}{3} \right)\Gamma_Q 
-{1 \over 45} k^{12}\lambda_{F}^{12}~\Gamma_Q^2 + \dots  \,. 
\end{equation}

This is a double expansion in powers of the parameters $\Gamma_Q$ and
$k\lambda_{F}$. The collisionless regime is in principle characterized 
by $\Gamma_Q \ll 1$, although, as was seen in Sec. 2, electron-electron interactions
can be neglected even when $\Gamma_Q \simeq 1$, as is the case for metals.
On the other hand, the fluid regime is characterized by small wave
numbers ($\Omega \ll \omega_p$). Indeed, keeping terms to fourth order in $k\lambda_{F}$,
Eq. (\ref{dispkin3}) reduces to the dispersion relation for
the effective SP system, Eq. (\ref{dispfluid2}). This is a further indication that
the effective SP system is a good approximation to the complete WP system
for long wavelengths.

We also note that for $\Gamma_Q \to 0$, the dispersion relation reduces to
\be
\omega^2 = \omega_{p}^2 + k^{2}v_{F}^2~.
\ee
This is exactly the dispersion relation obtained from the classical Vlasov-Poisson
system with a zero-temperature Fermi-Dirac equilibrium. In other words, when the
quantum coupling parameter is vanishingly small, a classical dynamical equation
can be used, as the only quantum effects come from the Fermi-Dirac statistics.
This situation may apply to extremely dense astrophysical systems 
such as white dwarfs.

\subsection{Stationary solutions}

As a second illustration, we use the present formalism to
describe the stationary states of the electron gas \cite{Haas}.
This result is more easily obtained by using the fluid version of our model.
In the time-independent case, the continuity equation (\ref{cont}) and 
the force equation (\ref{force2}) possess the following first integrals
\begin{equation}
\label{cons}
J = A^{2}u \,, \quad E = \frac{mu^2}{2} - e\phi + W -
\frac{\hbar^2}{2mA}\frac{d^{2}A}{dx^2} \,, 
\end{equation}
where $A = \sqrt{n}$. The first integrals in Eq. (\ref{cons}) 
corresponds to current ($J$) and energy ($E$) conservation. We can always choose $E = 0$ 
by a shift in the electrostatic potential. In this way, we can reduce 
the description of the stationary states to a set of second-order nonlinear 
ordinary differential equations for the amplitude $A$ and the electrostatic 
potential $\phi$. 
For a zero-temperature one-dimensional electron gas, 
the effective potential $W$ is given by Eq. (\ref{efpotdeg}); thus
from Eqs. (\ref{cons}), (\ref{po}) we get
\begin{eqnarray}
\label{stat1}
\hbar^2 \frac{d^{2}A}{dx^2} &=& m \left(\frac{mJ^2}{A^3} - 
2eA\phi + \frac{mv_{F}^{2}}{n_{0}^2}A^{5}\right) \,, \\
\label{stat2}
\frac{d^{2}\phi}{dx^2} &=& \frac{e}{\varepsilon_{0}}(A^2 - n_{0}) \,. 
\end{eqnarray}
Notice that, if the amplitude $A(x)$ is a slowly varying function of $x$, the
second derivative on the left hand-side of Eq. (\ref{stat1})
can be neglected. With this assumption, 
Eq. (\ref{stat1}) reduces to an algebraic equation, which
can be solved for $A$, and the result plugged into Eq. (\ref{stat2}).
This becomes a nonlinear differential equation for the electrostatic
potential, which is nothing but the Thomas-Fermi approximation to our model.

It can be easily verified that the $J = 0$ case cannot sustain small-amplitude, periodic
solutions. Hence, we assume $J = n_{0}u_0$ with $u_0 \neq 0$ and introduce the
following rescaling
\begin{eqnarray}
\label{resc1}
\hat{x} &=&  \frac{\omega_{p}x}{u_0}~ ,~~~
\hat{A} = \frac{A}{\sqrt{n_0}} ~,~~~ 
\hat{\phi} = \frac{e\phi}{m u_{0}^2} \nonumber\\
\label{resc2}
H &=& \frac{\hbar\omega_p}{m u_{0}^2}~ ,~~~
V_F = \frac{v_F}{u_{0}} \,.
\end{eqnarray}
We obtain, in the transformed variables 
(omitting the circumflex for simplicity of notation),
\begin{eqnarray}
\label{bgka}
H^{2}\frac{d^{2}A}{dx^2} &=& - 2\phi\,A + \frac{1}{A^3} + V_F^2 A^5\,, \\
\label{bgkphi}
\frac{d^{2}\phi}{dx^2} &=& A^2 - 1 \,,
\end{eqnarray}
a system that only depends on the rescaled parameters $H$ and $V_F$.
Note that the quantum coupling parameter can be written as
$\Gamma_Q = H/V_F^2$.

It is interesting to perform a linear stability analysis in order to see in
what conditions the system supports small amplitude spatially periodic
solutions. Writing
\begin{equation}
A = 1 + A'\exp(ikx) \,,\quad \phi = (1+V_F^2)/2 + \phi'\exp(ikx)~,
\end{equation}
and retaining in Eqs. (\ref{bgka})--(\ref{bgkphi}) only terms up to first order in
the primed variables, we obtain the relation
\begin{equation}
\label{quartic}
H^{2}k^{4} - 4 (1-V_F^2) k^2 + 4 = 0 \,.
\end{equation}
This second degree equation has solutions
\begin{equation}
k^{2} = \frac{2(1-V_F^2) \pm 2\sqrt{(1 - V_F^2)^2-H^{2}}}{H^2}
\,.
\end{equation}
Clearly, spatially oscillating solutions only exist when $k^2$ is
real and positive, which yields the condition
\be
V_F^2 < 1 - H~,
\ee
or equivalently
\be
m u_{0}^2 > m v_F^2 + \hbar \omega_p ~.
\ee
This expression sets a lower bound on the speed $u_0$,
below which no oscillating stationary solution can exist.

\subsection{Two-stream instability}

A classical plasma composed of two counterstreaming electronic populations 
with velocities $\pm u_0$ can give rise, for certain wave numbers, 
to an instability.
In a previous paper \cite{Haas}, we have shown that quantum effects
modify the dispersion relation, and give rise to a new instability branch.
These results were obtained by neglecting the effects of quantum statistics, 
and are therefore valid in the limit $v_F \ll u_0$.
Here, we perform the same calculations for finite values of $v_F$.	

We consider two electronic populations, which are both distributed according to
a zero-temperature Fermi-Dirac equilibrium, but with average velocities
$\pm u_0$. The motionless ions provide a neutralizing background.
The dispersion relation for such a two-stream plasma can be found in the
following way. For a single stream propagating at velocity $\pm u_0$,
our fluid model yields the following dielectric constant (thus valid for
long wavelengths) 
\be
\epsilon_\pm(k,\omega) = 1 - 
\frac{\omega_p^2}{(\omega \mp ku_0)^2 -k^2v_F^2 -\hbar^2k^4/4m^2}~.
\ee 

Setting $\epsilon_\pm(k,\omega) = 0$ leads to the dispersion 
relation found previously, Eq. (\ref{dispfluid}),
with the appropriate Doppler shift. The dielectric constant for the two-stream
case is found by averaging the contributions from each stream
$\epsilon(k,\omega) = (\epsilon_{+}~ + ~\epsilon_{-})/2$.
Using the normalization of Eqs. (\ref{resc1}), we obtain
\be
\label{eps2s}
\epsilon(k,\omega) = 1-\frac{1/2}{(\omega + k)^2 -k^2 V_F^2 -H^2k^4/4}
- \frac{1/2}{(\omega - k)^2 -k^2 V_F^2 -H^2k^4/4}~.
\ee

Setting $\epsilon(k,\omega) = 0$, we obtain the dispersion relation for the
two-stream plasma
\[
\omega^4 ~- ~\left(1 + 2k^2(1+V_F^2) + \frac{H^{2}k^{4}}{2}\right)\omega^2 
\]
\be
\label{disp2s}
- ~k^{2}\left(1-V_F^2 - \frac{H^{2}k^2}{4}\right)\left(1 - (1-V_F^2) k^2 + 
\frac{H^{2}k^4}{4}\right) = 0 \,.
\ee 

Notice that for $V_F = 0$ we recover the dispersion relation obtained
in \cite{Haas}.
Solving for $\omega^2$, one obtains
\be
\omega^2 = \frac{1}{2} + k^{2}\left(1 + v_{F}^2 + \frac{H^{2}k^2}{4}\right) 
\pm \frac{1}{2}\left[1 + 8k^{2}\left(1 + 2k^{2}V_{F}^2 + 
\frac{H^{2}k^4}{2}\right)\right]^{1/2} ~.
\ee

The solution for $\omega^2$ has two branches, one of which
is always positive and gives stable oscillations. The
other solution is negative ($\omega^2 < 0$) provided that

\begin{equation}
\label{cond}
[H^{2}k^{2} -4(1-V_F^2)][H^2 k^4 -4(1-V_F^2)k^2 + 4 ] < 0~.
\end{equation}

We immediately notice that, if $V_F \ge 1$, Eq. (\ref{cond}) is never verified,
and therefore there is no instability. This is a quite natural result. Indeed, 
mathematically, the instability is due to the fact that the two-stream
velocity distribution has a `hole' around $v=0$. When $V_F \ge 1$, the hole
is filled up, and no instability can occur. To put it differently, there can be
instability only when the equilibrium distribution 
is a non-monotonic function of the energy, which
ceases to be true when $V_F \ge 1$.

When $V_F <1$, Eq. (\ref{cond}) bifurcates for $H = 1-V_F^2$.
If $H \ge 1-V_F^2$, the second factor is always positive, and instability occurs
for $H^2 k^2 < 4(1-V_F^2)$. If $H < 1-V_F^2$, 
there is instability if either

\begin{equation}
\label{cond1}
0 < H^2 k^2 < 2(1-V_F^2)-2\sqrt{(1-V_F^2)^2-H^2} ~,
\end{equation}
or
\begin{equation}
\label{cond2}
2(1-V_F^2)+2\sqrt{(1-V_F^2)^2-H^2} < H^2 k^2 < 4(1-V_F^2)~.
\end{equation}

This yields the stability diagram plotted on Fig. 1, which is 
similar to the one obtained in the limiting case $V_F =0$.
The presence of a finite Fermi velocity has the effect of reducing 
the region of instability.
Numerical simulations yield similar reult to those observed in the
$V_F = 0$ case, which are reported in \cite{Haas}.


\section{Conclusion}

In this paper, we have first established the conditions of validity
of the Wigner-Poisson system. Subsequently, by taking moments
of the Wigner equation, we have derived an effective Schr\"odinger-Poisson system 
that captures the essential features of 
a quantum electron gas. In the long wavelength limit, 
this model correctly reproduces the results  
of the linear analysis of the Wigner-Poisson system. 
The advantages of the effective SP model are manifold: it is local in space
(compared to the nonlocal WP system); it is cast into the ordinary
space, rather than the phase space; and it has a straightforward interpretation
in terms of fluid quantities. Furthermore, it is easily amenable to numerical 
studies, given the abundance of accurate numerical techniques for the Schr\"odinger
equation (in comparison, numerical methods for 
the Wigner equation \cite{Suh} are much scarcer and more cumbersome
to implement).
The crucial points in
the derivation of the model are : (a) the decomposition of the pressure
into a classical and a quantum contribution, and (b)
the restriction to an appropriate class of statistical mixtures
(composed of states with same amplitude, but different phases). 
We believe that this class is wide enough to describe a significant
range of relevant physical systems.

For the case of a completely degenerate electron gas, 
the effective SP model can be put in a particularly simple form, in
which the Schr\"odinger equation exhibits a $\vline\Psi\vline~^4$ nonlinearity.
This model has been applied to the study of linear wave propagation,
nonlinear stationary solutions and the two-stream instability.
The simplicity of the resulting system of equations makes it a useful tool
for the study of quantum transport in solid state plasmas.


\newpage

{\large {\bf Acknowledgments}}

We would like to thank Pierre Bertrand for his valuable comments 
and suggestions. One of us (F. H.) thanks the Laboratoire de 
Physique des Milieux Ionis\'es for hospitality while this work 
was carried out and the Brazilian agency Conselho Nacional de Desenvolvimento
Cient\'{\i}fico e Tecn\'ologico (CNPq) for financial support. 


\newpage
\begin{center}
{\bf Figure Captions}
\end{center}

{\bf Figure 1} : Stability diagram for the two-stream plasma,
with  $V_F = 0.7$ (solid lines) and $V_F = 0$ (dashed lines). 
The curves correspond to Eqs. (\ref{cond1})-(\ref{cond2}).
For both cases, the region of the plane containing the $H^2$ axis is unstable.

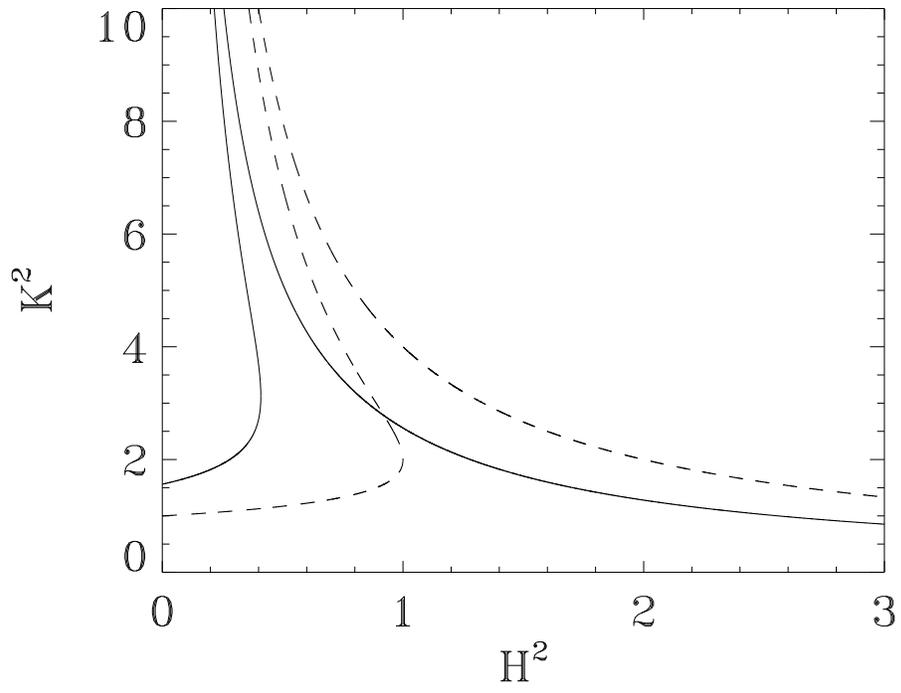
\end{document}